\documentstyle[12pt]{article}
\begin{document}

\title{
Isospin effect on nuclear stopping in intermediate
energy Heavy Ion Collisions \footnote{Supported by
National Natural Science Foundation of China under No. 19975073, National Science Foundation 
of Nuclear Industry under No. Y7100A0101 and Major State 
Basic Research Development Program in China under contract No. G20000774}}
\date{}
\maketitle

\begin{center}
\author{\normalsize Qingfeng Li$^{1)}$, Zhuxia Li$^{1,2,3)}$\\
\small 1) China Institute of Atomic Energy, P. O. Box 275 (18), 
Beijing 102413, P. R. China\\
2) Institute of Theoretical Physics, 
Academia Sinica, P. O. Box 2735, Beijing 100080, P. R. China\\
3) Center of Theoretical Nuclear Physics, National
Laboratory of Lanzhou Heavy Ion Accelerator, Lanzhou 730000, P. R. China
}
\end{center}

\begin{abstract}
By using the Isospin Dependent Quantum Molecular Dynamics Model (IQMD), 
 we study the dependence of nuclear stopping $Q_{ZZ}/A$ and $R$ in intermediate energy 
heavy ion collisions on system size, initial $N/Z$, isospin symmetry potential 
and the medium correction of two-body cross sections. 
We find the effect of initial $N/Z$ ratio, isospin symmetry potential
on stopping is weak. The excitation function of $Q_{ZZ}/A$ and $R$
depends on the form of medium correction of two-body cross sections, 
the equation of state of nuclear matter (EOS). Our results show the behavior of 
the excitation function of $Q_{ZZ}/A$ and $R$ can provide clearer information 
of the isospin dependence of the medium correction of 
two-body cross sections.   
\end{abstract}
\vspace{0.5cm}

It is important to investigate whether an equilibrium is reached or not 
for a colliding system in order to obtain the information of EOS and the 
reaction mechanism correctly. This problem has been studied extensively 
both theoretically and experimentally \cite{Cu81}-\cite{Li94}.
Following the establishment of  
radioactive beam facilities in many laboratory, it becomes possible to study
neutron (or proton) rich nuclear collisions at intermediate energies. 
Therefore it is necessary to study the effect of isospin asymmetry on the 
equilibrium process in a colliding system. 
In \cite{Liu01} the sensitivity of nuclear stopping to the isospin dependence of cross section 
was studied but 
the isospin dependence of the medium correction of two-body cross sections and the 
isospin asymmetry of the initial system was not tested. 
As we expect the degree of equilibrium is influenced 
by the mean field, medium correction of two-body cross section as well as the size of colliding system. 
To considering the influence of the isospin dependence of the medium correction of two-body cross section, 
in this letter we use 4 different forms of in-medium two-body cross sections, 
they are: 1) $\sigma_{0}$, the free nucleon-nucleon cross section 
which is isospin dependent \cite{Ch68}, 2) $\sigma_{0}^{*}$ \cite{Kl93}, which reads as
\begin{equation}
\sigma^{*}=[1+\alpha(\rho/\rho_{0})]\sigma^{free},  \label{eq1}
\end{equation}
here $\alpha=-0.2$. 3) $\sigma_{1}$ according to Ref.\ \cite{Li9394}. 4) $\sigma_{2}$, 
in which the medium correction is obtained based on the results of \cite{Li00}. 
In \cite{Li9394} G. Q. Li and R. Machleidt 
calculated the in-medium  neutron-proton, proton-proton (neutron-neutron) scattering cross section 
at energies up to 300 AMeV with the Dirac-Brueckner Hartree-Fock approach where the medium 
correction 
of $\rho$ meson mass was not taken into account and it was found that the in-medium cross sections 
decreased as density and  increased slightly when density was higher than normal nuclear density 
and energy was higher than about 125 AMeV (in Lab. system). 
In \cite{Li00} the in-medium binary scattering cross section is derived    
based on QHD-II type Lagrangian within the framework of the microscopic transport 
theory at energies up to 800 AMeV where the medium correction of $\rho$ meson mass
was taken into account \cite{Lolos} and it was found that because of the medium correction of 
$\rho$ meson mass 
the medium correction of nucleon-nucleon cross section was also 
isospin dependent, i.e., $\sigma_{np}^{*}$ depends on the baryon density weakly 
while $\sigma_{nn(pp)}^{*}$ depends on the baryon density obviously and a slightly
increasing with energy when density was higher than normal density and energy was higher 
than about 200 AMeV (in Lab. system) was also shown. 
The influence of the mean field is considered by taking EOS to be 'soft' EOS ($K=200 MeV$) and 
'hard' EOS ($K=380 MeV$) 
as well as with and without isospin symmetry potential. 
In order to  study the dependence of the degree of equilibrium on the nuclear size and the 
initial isospin asymmetry, we calculate two groups of colliding systems 1) $^{58}Ni+^{58}Ni$ and 
its neighbor nuclei with $A=60$ $Z=24-32$ and 2) $^{120}Sn+^{120}Sn$ and its neighbor nuclei
with $A=120$ $Z=47-55$. 

We study the degree of equilibrium in momentum space by investigating 
the momentum quadrupole $Q_{ZZ}$ and $R$, the ratio between perpendicular momentum 
and parallel momentum, which usually are called nuclear stopping. 
$Q_{ZZ}$ is defined as

\begin{equation}
Q_{ZZ}=\Sigma_{i}[2*P_{z}(i)^{2}-P_{x}(i)^{2}-P_{y}(i)^{2}],  \label{eq2}
\end{equation}
and $R$ is defined as
\begin{equation}
R=\frac{2}{\pi}(\Sigma_{i} |P_{\bot}(i)|)/(\Sigma_{i} |P_{z}(i)|).  \label{eq3}
\end{equation}
Here the summation runs over all nucleons in projectile and target and
the transverse momentum $P_{\bot}(i)=\sqrt{P_{x} (i)^{2}+P_{y} (i)^{2}}$.

The Isospin Dependent Quantum
Molecular Dynamics (QMD) model \cite{Ai91,Har89} is used in the calculations. 
For details, please see ref. \cite{Li01}. The used parameters of EOS are shown in 
Table 1.

First of all, we investigate the effect of symmetry potential and isospin dependence of the 
medium correction of two-body cross section as well as the isospin asymmetry of initial 
system on $Q_{ZZ}$. Fig. 1 shows average $Q_{ZZ}$ as function of Z for projectile 
and target taken to be a) $A=60$, $Z$ ranging 
from $24$ to $32$ and b) $A=120$, $Z$ ranging from $47$ to $55$ at beam energy $50 AMeV$, $150 AMeV$  
and $400 AMeV$ with $C_{S}=0$ and $35$ MeV
and two-body cross section taking to be $\sigma_{0}$ and $\sigma_{2}$, 
respectively.  As a whole, we can find that the 
effects of symmetry potential and the initial isospin asymmetry on $Q_{ZZ}$ are weak 
for both in-medium cross sections taken to be $\sigma_{0}$ 
(no medium correction) and $\sigma_{2}$ (with isospin dependent medium correction) cases. 

\[
\fbox{Fig. 1 a), b)} 
\]

In Figs. 2 a) and b) we present the excitation function of average $Q_{ZZ}$ with 
different EOS ('Soft' and 'Hard') and cross sections ($\sigma_{0}$, $\sigma_{0}^{*}$, 
$\sigma_{1}$  and $\sigma_{2}$) for  a) $^{58}Ni+^{58}Ni$ and b) 
 $^{120}Sn+^{120}Sn$. 
It is shown that $Q_{ZZ}/A$ first increases with beam energy then it is saturated at   
about 150-300 AMeV depending on the in-medium cross section used. 
Different forms of medium correction of two-body cross sections significantly 
affect the behavior of excitation function of $Q_{ZZ}/A$. 
From comparison between Fig. 2 a) and b) we find that for $^{58}Ni+^{58}Ni$ case
the influence of different EOS on $Q_{ZZ}/A$ is much weaker than that of different form of 
cross sections while for  $^{120}Sn+^{120}Sn$ 
case it is comparable with that of different cross sections. 
So from above study we may conclude that the behavior of the excitation function of  
$Q_{ZZ}/A$ for medium size system like  $^{58}Ni+^{58}Ni$ can provide clearer 
information
of isospin dependence of the medium correction of cross sections. 

\[
\fbox{Fig. 2 a), b)} 
\]

Figs. 3 a) and b) show the excitation function of $R$ for a) $^{58}Ni+^{58}Ni$ and b) 
$^{120}Sn+^{120}Sn$ with different EOS ('Soft' and 'Hard'), cross section  
($\sigma_{0}$, $\sigma_{0}^{*}$, and $\sigma_{2}$). For both $^{58}Ni+^{58}Ni$ and 
$^{120}Sn+^{120}Sn$ cases, $R$ decreases when beam energy is less than $\sim 100 AMeV$ 
and there is a change of the decreasing slope at 50 AMeV. 
As beam energy further increases $R$ increases. For $^{58}Ni+^{58}Ni$ case, the slope of 
excitation function of $R$ is dominated by the form of in-medium 
cross sections and its dependence on the mean field is very weak. 
While for $^{120}Sn+^{120}Sn$ case, the slope of excitation function of  
$R$ depends on both form of medium correction of
two-body cross section and the EOS of nuclear matter. This behavior of $R$ depending on   
system size is the same as $Q_{ZZ}/A$ case shown in Figs. 2 a) and b).  It implies that
the measurement of the excitation function of the stopping in medium size system like 
$^{58}Ni+^{58}Ni$ can provide 
clearer information of medium correction of two-body cross section.

\[
\fbox{Fig. 3 a), b)} 
\]

Fig. 4 shows the excitation function of $R$ for $^{120}Sn+^{120}Sn$ at different impact 
parameters. The behavior of $R$ as function of energy changes quickly for different 
impact parameters. The slopes of $R$ as function of beam energy change 
sign from negative to positive at $E\sim 100 AMeV$ for $b=0 fm$ - $b=3 fm$ cases. 
For low beam energy the slope of excitation function depends on impact parameter weakly. 
It implies that for lower energy case the stopping does not depend on impact parameter strongly
but as energy increases the stopping does depend on impact parameter strongly. 

\[
\fbox{Fig. 4} 
\]

In summary, in this paper we study the dependence of nuclear stopping $Q_{ZZ}/A$ and $R$  
on isospin asymmetry of the initial system, the isospin dependence of the medium correction
of two-body scattering cross section and EOS of nuclear matter (isospin dependent part and 
incompressibility). 
We find that the dependence of stopping on initial isospin asymmetry and isospin dependent part 
of nuclear potential 
is weak and the dependence on the different forms of the medium correction of two-body cross 
section is strong. By studying the excitation function of $Q_{ZZ}/A$ and $R$ we find the
influence of difference of EOS (incompressibility) on stopping is comparable  
with that of different medium correction of two-body cross sections for $^{120}Sn+^{120}Sn$ 
but is much weak for $^{58}Ni+^{58}Ni$ case at energy range studied in this work. Therefore 
the measurement of the excitation function of nuclear stopping for collisions of medium size 
nuclei is a better means to extract 
the information of medium correction of two-body cross section. We also find that 
the dependence of nuclear stopping on impact parameter is weak at low energy and it becomes 
stronger at energy higher than $\sim 50-100 AMeV$.

\begin{description}
\item[{\tt Table. 1}]  Parameters used in calculations
\end{description}

\begin{tabular}{|l|c|c|c|c|c|c|c|}
\hline
EOS&$\alpha (MeV)$ & $\beta (MeV)$ & $\gamma $ & $\rho _0 (fm^{-3})$ & $K (MeV)$ & 
$L (fm)$ & $C_{Yuk} (MeV)$ \\ \hline
Soft&$-356$ & $303$ & $7./6.$ & $0.168$ & $200$ & $2.1$ & $-5.5$ \\ \hline
Hard&$-124$ & $70.5$ & $2.$ & $0.168$ & $380$ & $2.1$ & $-5.5$ \\ \hline
\end{tabular}

\newpage
\begin{center}
{\bf CAPTIONS}
\end{center}
\begin{description}
\item[{\tt Fig. 1}]  The average $Q_{ZZ}$ as function of Z  for projectile 
and target taken to be a) $A=60$ and $Z$ ranging from $24$ to $32$ and 
b) $A=120$, $Z$ ranging from $47$ to $55$ at beam energy $50 AMeV$, $150 AMeV$ and  
$400 AMeV$ with $C_S=0$ and $35$ MeV
and two-body cross section taking to be $\sigma_{0}$ and 
$\sigma_{2}$ (see text), respectively. 
\item[{\tt Fig. 2}]  The average $Q_{ZZ}$ as function of beam energy for a) $^{58}Ni+^{58}Ni$  
and b) $^{120}Sn+^{120}Sn$ at $b=0 fm$ with different EOS and two-body cross sections 
$\sigma$ (see text). 
\item[{\tt Fig. 3}] $R$ as function of beam energy at different EOS and two-body cross sections 
$\sigma$ (see text).  
\item[{\tt Fig. 4}] $R$ as function of beam energy with different input parameters $b$. 
\end{description}
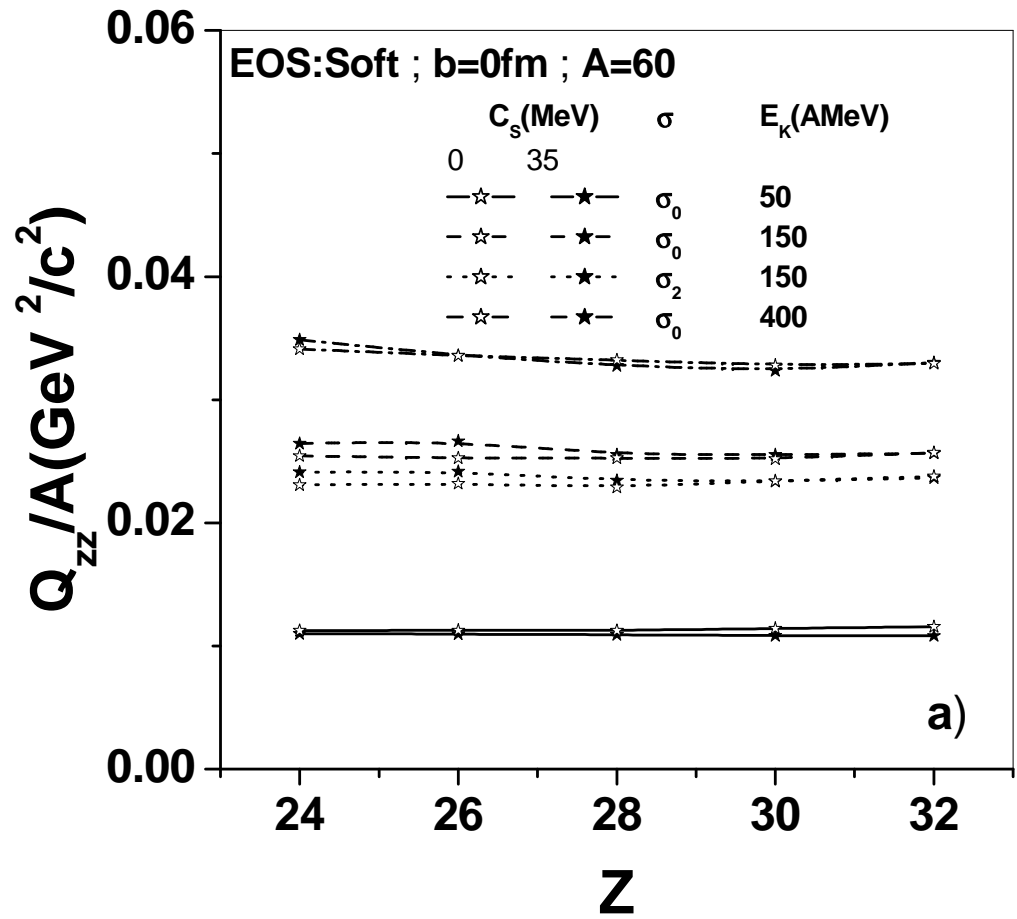
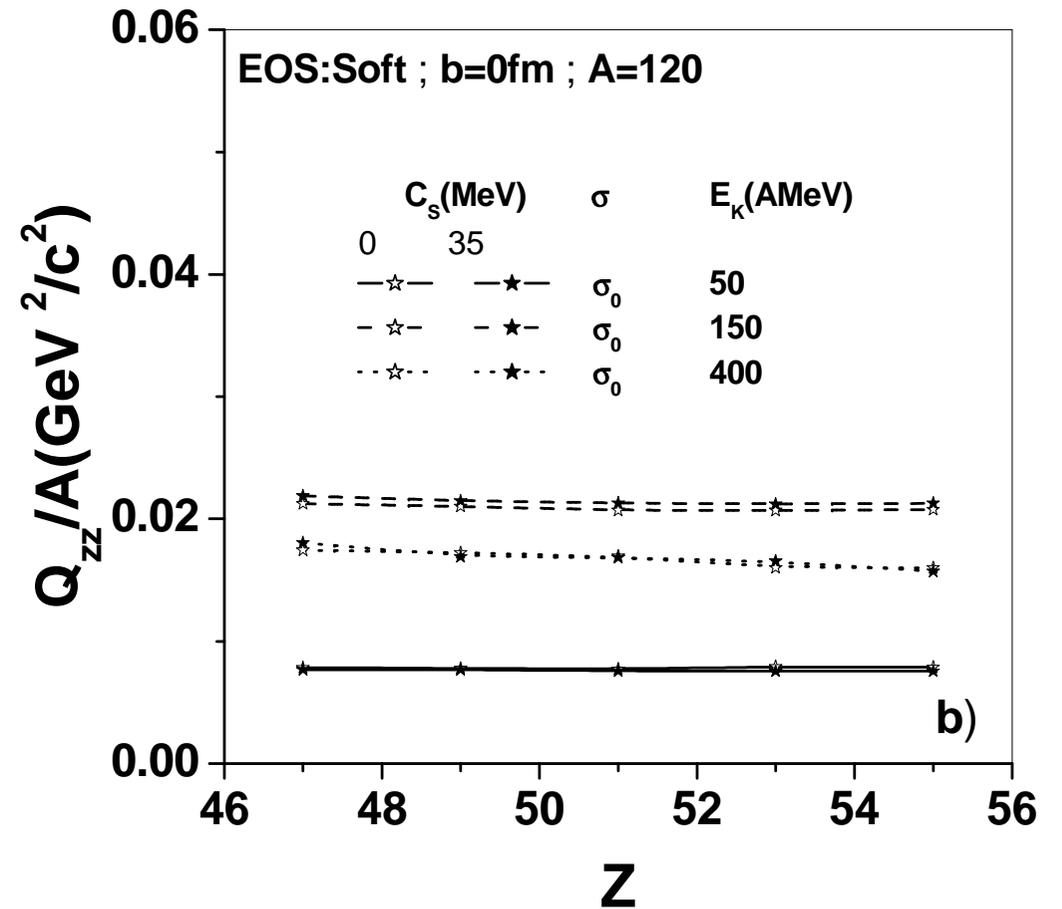
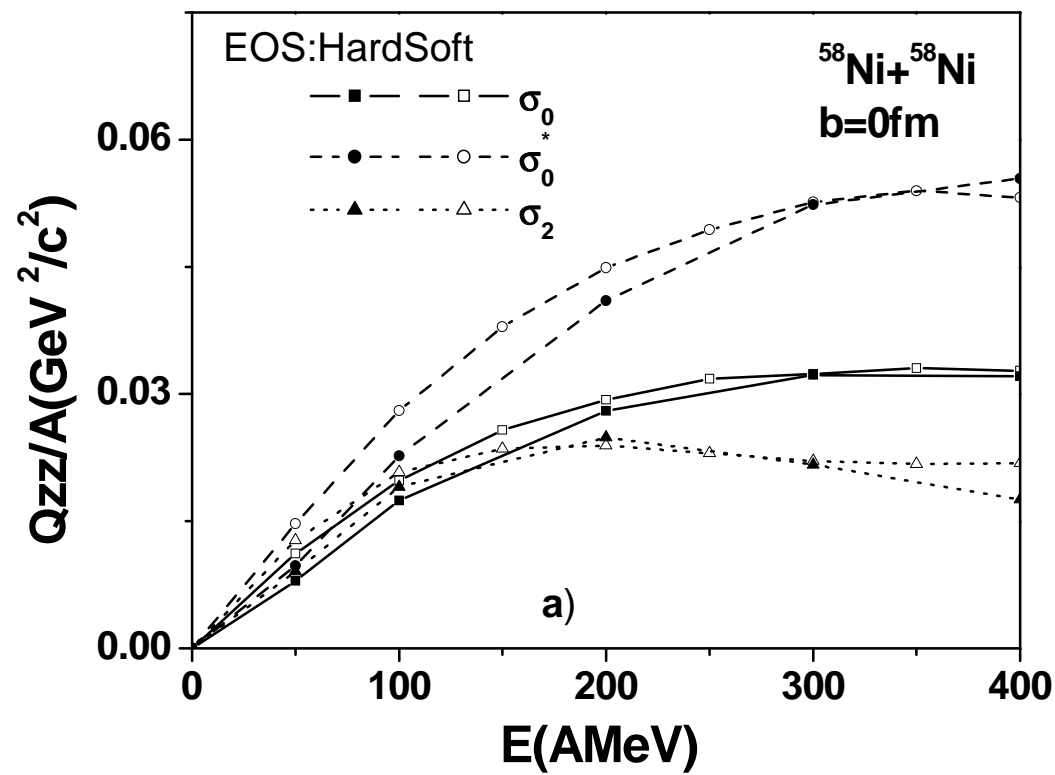
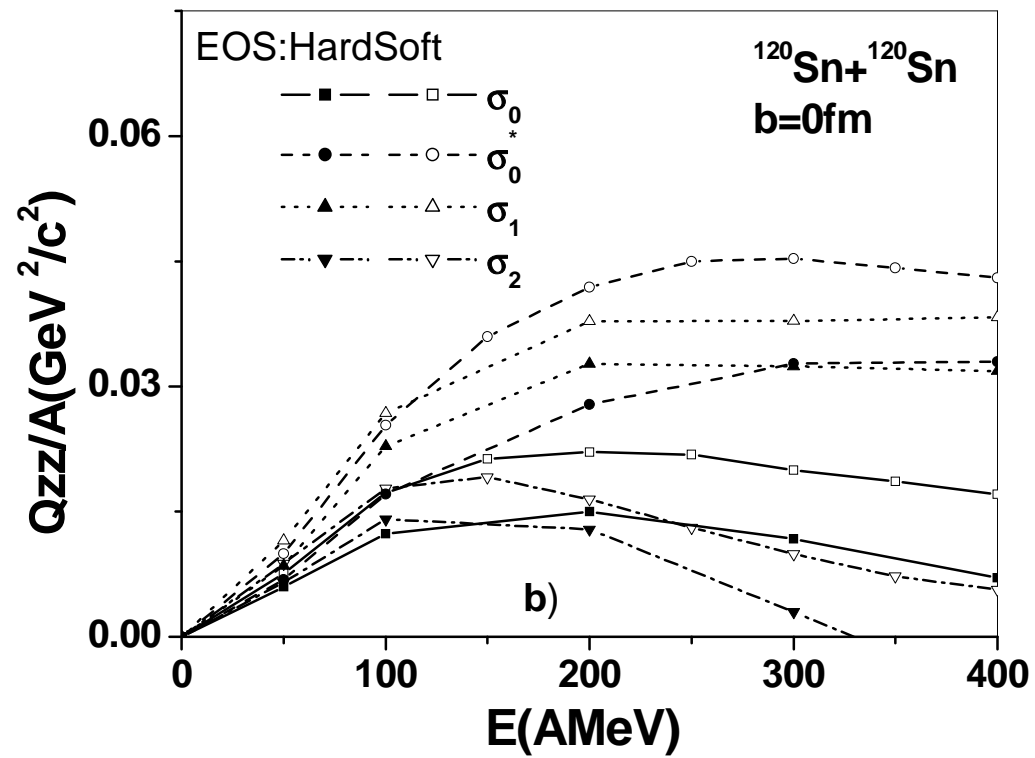
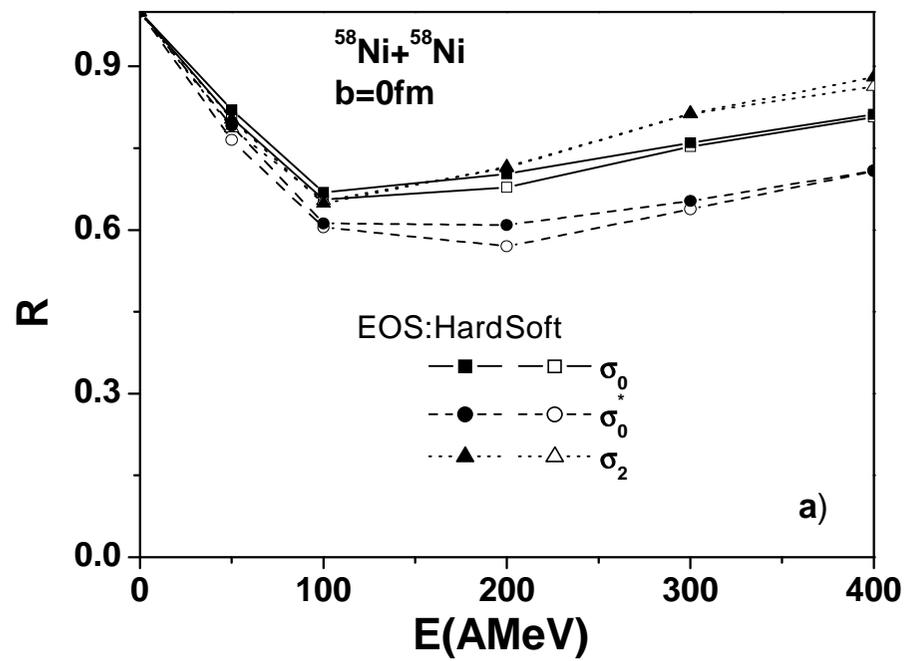
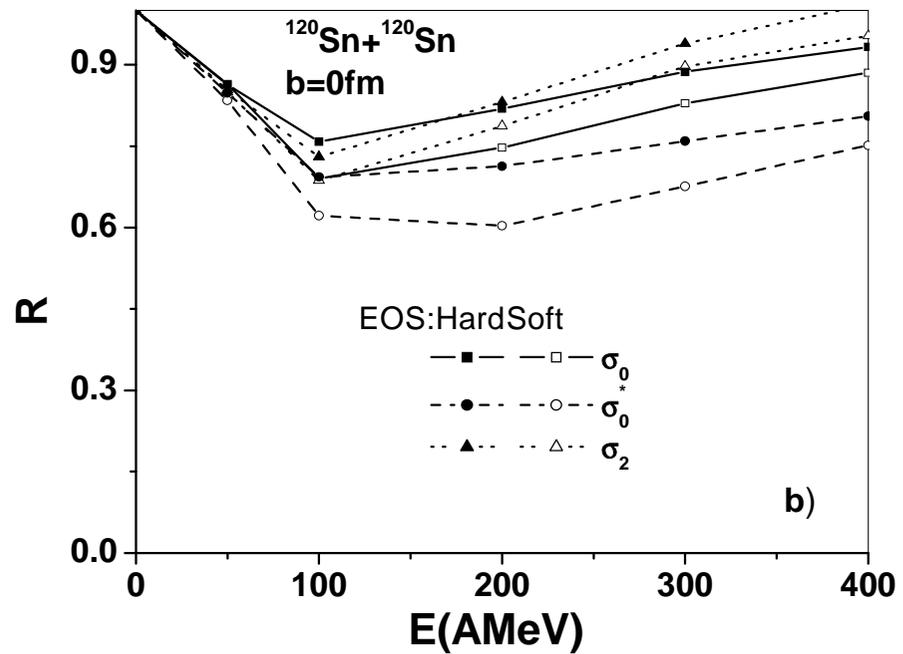
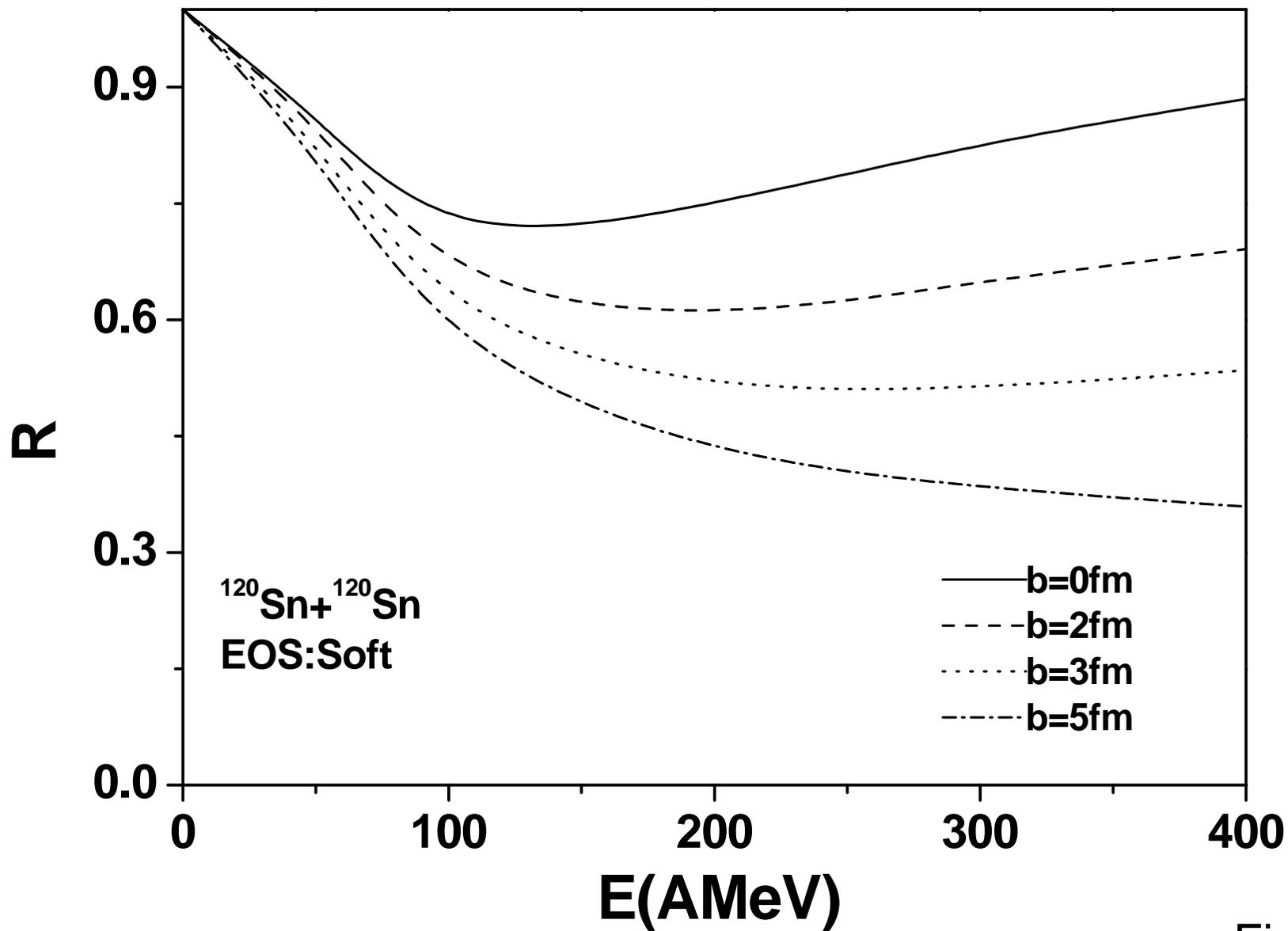
\end{document}